# (Ca,Na)(Zn,Mn)$_2$As$_2$: a new spin & charge doping decoupled diluted ferromagnetic semiconductor with hexagonal CaAl$_2$Si$_2$ structure

K. Zhao[1], B.J. Chen[1], Z. Deng[1], W. Han[1,6], G.Q. Zhao[1], J. L. Zhu[1], Q. Q. Liu[1], X. C. Wang[1], B. Frandsen[2], L. Liu[2], S. Cheung[2], F. L. Ning[3], T.J.S. Munsie[4], T. Medina[4], G.M. Luke[4], J.P. Carlo[5], J. Munevar[7], G. M. Zhang[8], Y. J. Uemura[2] and C. Q. Jin[1,2]*

1. Beijing National Laboratory for Condensed Matter Physics and Institute of Physics, Chinese Academy of Sciences, Collaborative Innovation Center of Quantum Matter, Beijing, China

2. Department of Physics, Columbia University, New York, New York 10027, USA

3. Department of Physics, Zhejiang University, Hangzhou 310027, China

4. Department of Physics & Astronomy, McMaster University, Hamilton, Canada.

5. Department of Physics, Villanova University, Villanova, PA 19085, USA

6. Department of Physics, Chengde Mingzu College, Chengde, China

7. Centro Brasileiro de Pesquisas Fisicas, Rio de Janeiro, Brazil

8. Department of Physics, Tsinghua University, Beijing, China

## Abstract

Here we report the successful synthesis of a spin- & charge-decoupled diluted magnetic semiconductor (Ca,Na)(Zn,Mn)$_2$As$_2$, crystallizing into the hexagonal CaAl$_2$Si$_2$ structure. The compound shows a ferromagnetic transition with a Curie temperature up to 33 K with 10% Na doping, which gives rise to carrier density of $n_p \sim 10^{20}$ cm$^{-3}$. The new DMS is a soft magnetic material with $H_C$<400 Oe. The anomalous Hall effect is observed below the ferromagnetic ordering temperature. With increasing Mn doping, ferromagnetic order is accompanied by an interaction between the local spin and mobile charge, giving rise to a minimum in resistivity at low temperatures and localizing the conduction electrons. The system provides an ideal platform for studying the interaction of the local spins and conduction electrons.

Diluted magnetic semiconductors (DMS) have received much attention due to their potential for application in the field of spintronics [1-5]. In typical III-V DMS systems, such as (Ga,Mn)As, (In,Mn)As and (Ga,Mn)N, substitution of divalent Mn atoms into trivalent Ga or In sites leads to severely limited chemical solubility, resulting in metastable specimens only available as epitaxial thin films. Moreover, the hetero-valence substitution, which simultaneously dopes both charge and spin, makes it difficult to individually control each quantum degree of freedom.

Recently, a new type of DMS, Li(Zn,Mn)As (termed "111" following the three chemical composition ratios), was discovered by Deng et al,[6] with charges injected via off-stoichiometry of Li concentrations and spins via isovalent ($Zn^{2+}$,$Mn^{2+}$) substitution, showing a Curie temperature ($T_C$) up to 50 K. Shortly thereafter, a new ferromagnetic DMS (Ba,K)(Zn,Mn)$_2$As$_2$ (named "122" following the terminology of iron pnictide superconductors[7]) was synthesized.[8] Exploiting (Ba,K) substitution to introduce hole carriers and (Zn,Mn) substitution to supply magnetic moments, compositions with 5-15 % Mn doping exhibit a ferromagnetic transition temperature up to 180 K.[7] The "122" DMS (Ba,K)(Zn,Mn)$_2$As$_2$ shares the tetragonal ThCr$_2$Si$_2$ structure with the "122" iron pnictide superconductor (Ba,K)Fe$_2$As$_2$ and the antiferromagnet BaMn$_2$As$_2$[9], each with a lattice mismatch of less than 2%, giving rise to the unprecedented possibility of designing various interface-based devices among DMS, superconductors and magnets. Moreover, an increase in the carrier density could enhance the Curie temperature further to 230K in this 122 system.[10] The idea of decoupling spin and charge has since been extended, resulting in the discovery of two additional DMS system, "111" type Li(Zn,Mn)P[11] and "1111" types (La,Ca)(Zn,Mn)SbO & (La,Ca)(Zn,Mn)AsO[12,13]

Depending on the composition, "122" compounds can crystallize into the tetragonal $ThCr_2Si_2$ structure, as in the case of $(Ba,K)(Zn,Mn)_2As_2$, or the hexagonal $CaAl_2Si_2$ structure. Interestingly, the hexagonal compounds $CaMn_2As_2$ and $CaMn_2Sb_2$ are fully frustrated classical magnetic systems with a honeycomb lattice of Mn atoms, exhibiting a complex magnetic phase diagram.[14] Here we report the synthesis of bulk specimens of a new DMS $(Ca,Na)(Zn,Mn)_2As_2$ with the hexagonal $CaAl_2Si_2$ structure, as shown in Fig. 1(a).[15] Hole carriers are provided through (Ca,Na) substitution and local spins through (Zn,Mn), resulting in ferromagnetism with Tc up to 33 K. Clear signatures of the ferromagnetic order are seen in the negative magnetoresistance and anomalous Hall effect below Tc. With increasing the Mn doping, the interaction between the local spins and conduction electrons gives rise to the minimum in resistivity and eventual localization of the conduction electrons at low temperatures.

**Experimental**

Polycrystalline samples of $(Ca,Na)(Zn,Mn)_2As_2$ were synthesized via the solid-state reaction method. The synthesis of starting materials CaAs and $Na_3As$ was described in an earlier paper.[8] These starting materials were mixed with high-purity Zn, Mn, and As powders according to the nominal composition of $(Ca,Na)(Zn,Mn)_2As_2$. The mixture was sealed inside an evacuated titanium tube that was, in turn, sealed inside an evacuated quartz tube. The mixture was heated to 750 °C at a rate of 3 °C/min. This temperature was maintained for 20 h before being slowly decreased to room temperature at a rate of 2 °C/min. Polycrystals were annealed at 600 °C for 10 h to remove the effect of grain boundaries in the transport measurements. Samples were characterized via X-ray powder diffraction with a Philips X'pert diffractometer using Cu K-edge radiation. The DC magnetic susceptibility was characterized by a superconducting quantum interference device magnetometer (Quantum Design, Inc.), and the electronic transport measurements were conducted with a physical property measuring system. Positive muon spin relaxation (μSR) measurements were performed on polycrystalline specimens at TRIUMF in Vancouver, Canada.

**Results and Discussion**

Fig. 1(b) shows the X-ray diffraction patterns of $(Ca_{0.9}Na_{0.1})(Zn_{1-x}Mn_x)_2As_2$ for $x = 0$, 0.05, 0.1, 0.15, 0.2, and 0.25, respectively, collected over a 2θ range from 10° to 80°. Rietveld refinements were performed to determine the symmetry and lattice parameters of all samples, as shown in Fig. 1(c). Compared with the lattice parameters $a$=4.1596Å and $c$=7.022Å for $(Ca_{0.9}Na_{0.1})Zn_2As_2$, the $c$-axis increases with doping concentration $x$, indicating successful solid solutions of Na and Mn.

Fig. 2(a) shows the temperature dependence of the magnetization in zero field cooling (ZFC) and field cooling (FC) procedures under 500 Oe for $(Ca_{0.9}Na_{0.1})(Zn_{1-x}Mn_x)_2As_2$ polycrystals with $x = 0.025$, 0.05, 0.1, 0.15, 0.2, and 0.25. The ferromagnetic transition temperature (Tc) initially increases with Mn doping, reaching a maximum value of about 33 K with 25% Mn content. As shown in the inset of Fig. 2(a), the sample $(Ca_{0.9}Na_{0.1})(Zn_{0.95}Mn_{0.05})_2As_2$ shows paramagnetic behavior at high temperatures with an effective paramagnetic moment of ~ 5 $\mu_B$ per $Mn^{2+}$ between

200 K and 300 K.

We also performed magnetic hysteresis measurements. Fig.2(b) shows the M(H) curves of $(Ca_{0.9}Na_{0.1})(Zn_{1-x}Mn_x)_2As_2$ at T = 2 K. The maximum saturation moment is 1.9 $\mu_B$ per Mn atom, comparable to that of $(Ga,Mn)As$[1], $Li(Zn,Mn)As$[6] and $(Ba,K)(Zn,Mn)_2As_2$[8]. From the inset of Fig 2(b), the coercive force increases with Mn doping and reaches a maximum of $H_C$<400 Oe, while the saturation moment mainly exhibits a decreasing tendency with Mn doping.

To examine the magnetically ordered volume fraction and the ordered moment size, μSR measurements were performed on a specimen of $(Ca_{0.9}Na_{0.1})(Zn_{0.95}Mn_{0.05})_2As_2$. Figure 2(c) shows the zero-field (ZF) μSR time spectra. A sharp increase of the muon spin relaxation rate is seen below 18 K, comparable to the $T_C$~20 K determined by susceptibility measurements (Fig. 2d). In Fig. 2(e), the temperature dependence of the volume fraction of the magnetically ordered state estimated from the μSR data is seen to be consistent with the spontaneous magnetization under 5 Oe, showing homogeneous ferromagnetism with $T^{3/2}$ dependence in low temperature.[16,17] However, it should be noted that the $T^{3/2}$ dependence in this case is related to the change in the volume fraction of the magnetically ordered phase, rather than the usual $T^{3/2}$ dependence arising from quadratic spin-wave dispersion in 100% magnetically ordered systems.

As shown in Table 1, the tradeoff between coercive force and saturation moment may be a result of competition between antiferromagnetic coupling of Mn moments in the nearest neighbor Zn sites and ferromagnetic coupling between Mn moments in more distant locations mediated by the doped hole carriers. As observed in $Bi_{2-x}Mn_xTe_3$[18], the saturation moments are much smaller than the effective paramagnetic moment above Tc, reminiscent of itinerant ferromagnetic systems[19]. The reason is probably the weak ferromagnetic exchange energy mediated by hole carriers between local Mn atoms through an RKKY-like interaction.

Resistivity measurements shown in Fig. 3(a) indicate that $CaZn_2As_2$ is a semiconductor[14]. Doping Na atoms into Ca sites introduces hole carriers, leading to metallic behavior in $(Ca,Na)Zn_2As_2$. Upon subsequent doping of magnetic moments, the system shows ferromagnetic order at low temperature, as observed from the resistivity curves of $(Ca_{0.9}Na_{0.1})(Zn_{1-x}Mn_x)_2As_2$ for selected values of $x$ up to 0.25 in Fig. 3(b). The scattering of carriers by spins provided by Mn dopants increases the resistivity at room temperature.

The resistivity of $(Ca_{0.9}Na_{0.1})(Zn_{0.95}Mn_{0.05})_2As_2$, displayed in Fig.3(c), decreases linearly as the temperature is lowered from high-temperature regime, similar to many metals. In the inset of Fig. 3(c), between 60 K and 70 K, the resistivity begins to deviate from the linear trend, presumably owing to the scattering caused by short-range ferromagnetic interactions, evidenced by the negative magnetoresistance described below. The decrease of lattice scattering and increase of magnetic scattering from short range ferromagnetic interactions leads to a local minimum in the resistivity at about 50 K. Around Tc, strong scattering from the ferromagnetic fluctuations causes a local maximum in the resistivity curve. Below Tc, owing to the establishment of ferromagnetic order and consequent decrease of magnetic scattering, the resistivity exhibits a rapid decline in low temperature region, consistent with the case of (Ga,Mn)As[1] and Li(Zn,Mn)As[6].

The magnetoresistance curve of $(Ca_{0.9}Na_{0.1})(Zn_{0.95}Mn_{0.05})_2As_2$ is depicted in Fig. 3(d) for temperatures between 2 K and 60 K. Similar to the case of metallic (Ga,Mn)As[20,21], obvious negative magnetoresistance develops for $T>T_C$, with the establishment of the short-range ferromagnetic interaction (without long range order). As the temperature decreases, the negative magnetoresistance is enhanced, and reaches its maximum near Tc probably due to the rapid polarization of the Mn spins in the ferromagnetic ordered state. At T = 2 K, an obvious hysteresis is observed in magnetoresistance curve, showing consistent behavior with the coercive force observed in the M(H) curve. The small positive magnetoresistance observed in low magnetic field is most probably caused by the rotation of spins from their original direction to the magnetic field direction.[21] The sample with $x = 0.1$ exhibits behavior similar to that described above and has a ferromagnetic Curie temperature of 28 K.

For the samples with $x = 0.15$ and $x = 0.2$, there exists an additional minimum in the resistivity at about 12K. As seen in Fig. 3(e), which displays the resistivity versus temperature on a logarithmic scale, the resistivity minimum is gradually suppressed as the applied magnetic field is increased, accompanying a type of antiferromagnetic interaction. On the semi-logarithmic scale, the resistivity is almost linear at low temperatures under different magnetic fields. It would be interesting for further study to investigate whether this phenomenon is the signature of the Kondo effect[22] or if it should be explained in terms of quantum corrections to the conductivity in the weakly localized regime for the spin-polarized universality class in (Ga,Mn)As[23]. The specimens with $x = 0.15$ exhibit similar transport behavior under an applied magnetic field as that of the $x = 0.2$ sample.

As the Mn concentration is increased to $x = 0.25$ in Fig. 3(f), the resistivity exhibits an increase at low temperature, implying the localization of conduction electrons, just as in the case of (Ga,Mn)N[20] and $(Ba,K)(Zn,Mn)_2As_2$[8]. The resistivity shows an anomaly at the ferromagnetic ordering temperature around 33 K, evidenced by the negative magnetoresistance below Tc.

To further investigate the transport properties, we measured the Hall effect for selected samples. The compounds with 10% Na doping show an anomalous Hall effect below the ferromagnetic ordering temperature with a carrier density of $n_p \sim 10^{20} cm^{-3}$, comparable to that of (Ga,Mn)As[1], (In,Mn)As[2] and Li(Zn,Mn)As[6]. The Hall resistivity of $(Ca_{0.9}Na_{0.1})(Zn_{0.85}Mn_{0.15})_2As_2$ at T=2 K, shown in Figure 4, reveals a coercive force of about 270 Oe, quite consistent with the M(H) curve. Considering how the transport properties can thus be effectively tuned by the ferromagnetic order, this new DMS is suitable for spin manipulation as a soft material in future.

## Conclusion

$(Ca,Na)(Zn,Mn)_2As_2$, a new spin- and charge-decoupled DMS based on the hexagonal $CaAl_2Si_2$-type structure, has been synthesized and characterized by structural, magnetic, μSR, and electronic-transport measurements. With (Ca,Na) substitution to introduce hole carriers and (Zn,Mn) substitution to introduce local spins, the system exhibits ferromagnetic order, which is evidenced by both μSR measurements and the anomalous Hall effect. With increasing Mn doping, the interaction of the local spins and conduction electrons gives rise to a minimum in resistivity at low temperature, ultimately localizing the conduction electrons. Thus, the system provides an ideal platform for research into the interaction of the local spins and conduction electrons.

## Acknowledgments:

The work was supported by NSF & MOST of China through research projects; the US NSF PIRE (Partnership for International Research and Education: OISE-0968226) and DMR-1105961 projects at Columbia; the JAEA Reimei project at IOP, Columbia, PSI, McMaster and TU Munich; and NSERC and CIFAR at McMaster.

Table 1: Th transition temperature $T_C$, saturation moment $M_S(H=0)$, and coercive field $H_C$ at T = 2 K after training in the external field of 1 T for $(Ca_{0.9}Na_{0.1})(Zn_{1-x}Mn_x)_2As_2$ with several different Mn spin doping levels $x$.

| $(Ca_{0.9}Na_{0.1})(Zn_{1-x}Mn_x)_2As_2$ | x=0.025 | x=0.05 | x=0.10 | x=0.15 | x=0.20 | x=0.25 |
|---|---|---|---|---|---|---|
| $T_C$[K] | 7 | 20 | 28 | 30 | 30 | 33 |
| Saturation Moment/ Mn | 1.2 | 1.9 | 1.0 | 0.7 | 0.45 | 0.38 |
| Coercive field/Oe | 42 | 105 | 165 | 270 | 260 | 400 |

**Figure Captions:**

**Fig. 1(a)** Crystal structure of $CaZn_2As_2$, with hexagonal $CaAl_2Si_2$ structure; **(b)** X-ray diffraction patterns of $(Ca_{0.9}Na_{0.1})(Zn_{1-x}Mn_x)_2As_2$ polycrystals for $x$ = 0, 0.05, 0.1, 0.15, 0.2, and 0.25, respectively; **(c)** $c$ axis lattice parameter of $(Ca_{0.9}Na_{0.1})(Zn_{1-x}Mn_x)_2As_2$.

**Fig. 2(a)** dc magnetization measured in H = 500 G in $(Ca_{0.9}Na_{0.1})(Zn_{1-x}Mn_x)_2As_2$ with several different charge doping levels $x$, with zero-field cooling (ZFC) and field-cooling (FC) procedures; **(b)** M(H) curves after field training up to 7 T and subtraction of the paramagnetic component measured in $(Ca_{0.9}Na_{0.1})(Zn_{1-x}Mn_x)_2As_2$ with several different charge doping levels $x$; **(c)** Zero-field μSR time spectra obtained from a polycrystalline specimen of $(Ca_{0.9}Na_{0.1})(Zn_{0.9}Mn_{0.1})_2As_2$; **(d)** the relaxation rate $a$ of the signal that exhibits fast relaxation; **(e)** Temperature dependence of the volume fraction of regions with static magnetic order, estimated by μSR measurements in zero field (ZF), consistent with that of spontaneous magnetization under 5 Oe.

**Fig. 3(a)** Temperature-dependent resistivity curves of $(Ca_{1-x}Na_x)Zn_2As_2$ for $x$ = 0, 0.05, and 0.1, respectively; **(b)** Resistivity of $(Ca_{0.9}Na_{0.1})(Zn_{1-x}Mn_x)_2As_2$ for $x$ = 0.1, 0.15, 0.2, 0.25, and 0.3, respectively; **(c)** Resistivity curve ρ(T) of $(Ca_{0.9}Na_{0.1})(Zn_{0.95}Mn_{0.05})_2As_2$, with dρ(T)/dT curve in the inset; **(d)** magnetoresistance curve of $(Ca_{0.9}Na_{0.1})(Zn_{0.95}Mn_{0.05})_2As_2$ measured in an external field of 7 T at T=2K, 5K, 10K, 20K, 30K, and 50K, respectively. Inset: magnetoresistance curve in low magnetic field at T=2K; **(e)** Resistivity curve ρ(T) of $(Ca_{0.9}Na_{0.1})(Zn_{1-x}Mn_x)_2As_2$ with $x$ = 0.2 under H=0T, 0.2T, 1T, and 2T, respectively. The temperature axis is on a logarithmic scale. There exists an additional minimum in the resistivity, and the low-temperature region is almost linear for various magnetic fields; **(f)** $\rho(T)$ of $(Ca_{0.9}Na_{0.1})(Zn_{1-x}Mn_x)_2As_2$ with $x = 0.25$ under H=0T, 0.2T, and 1T, respectively.

**Fig. 4** Hall effect results from a sintered specimen of $(Ca_{0.9}Na_{0.1})(Zn_{1-x}Mn_x)_2As_2$ with Mn doping level $x = 0.15$ at T = 2 K. Anomalous Hall effect and a very small coercive field are seen, consistent with the magnetic hysteresis curve M(H) at T = 2 K.

**Fig.1**

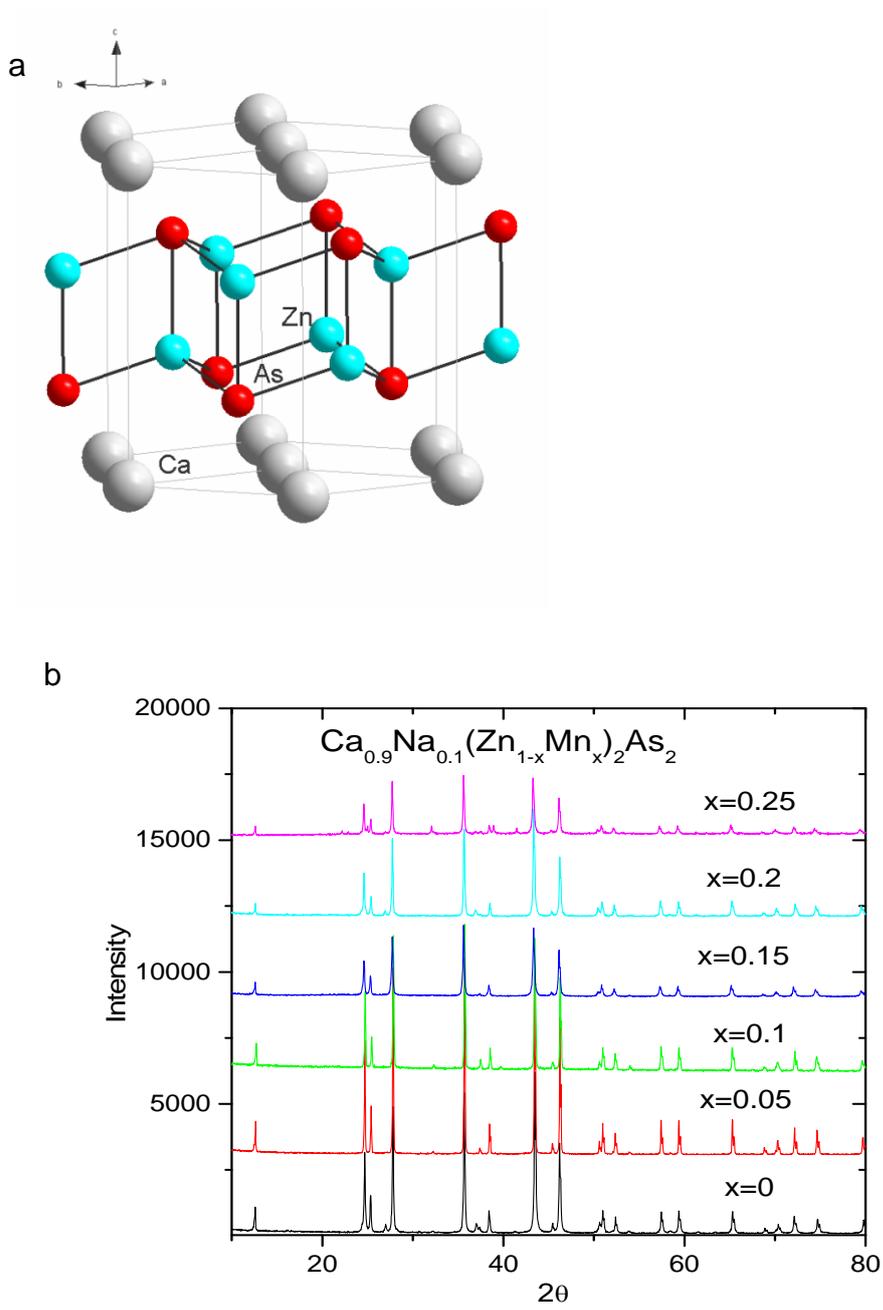

c

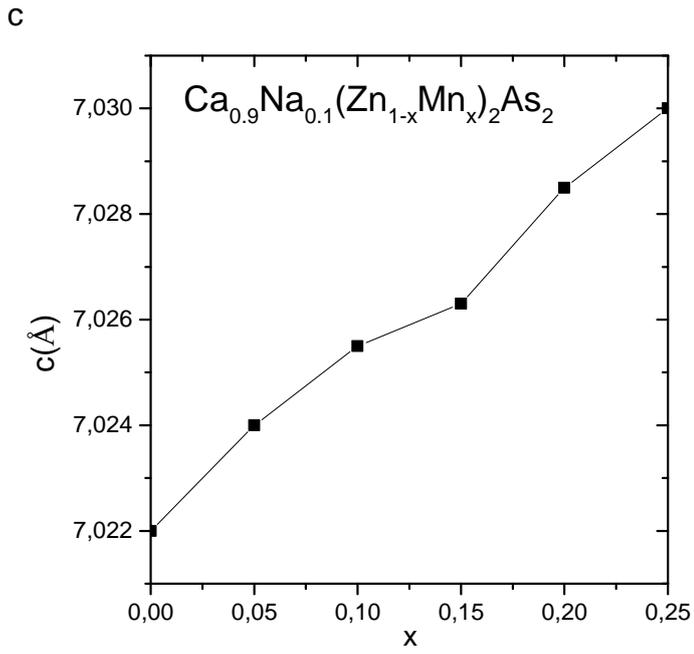

**Fig. 2**

a

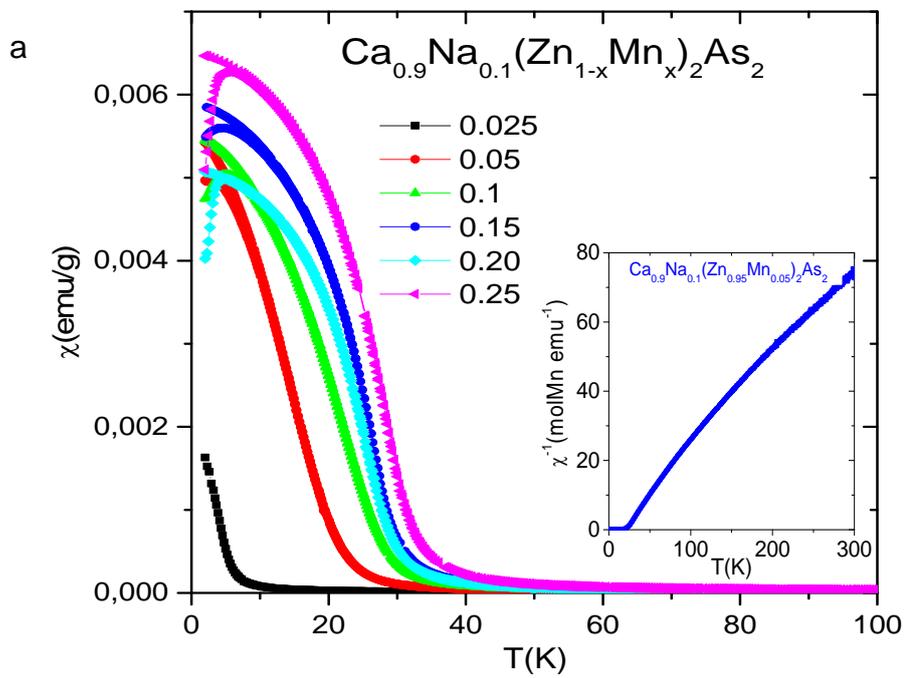

b

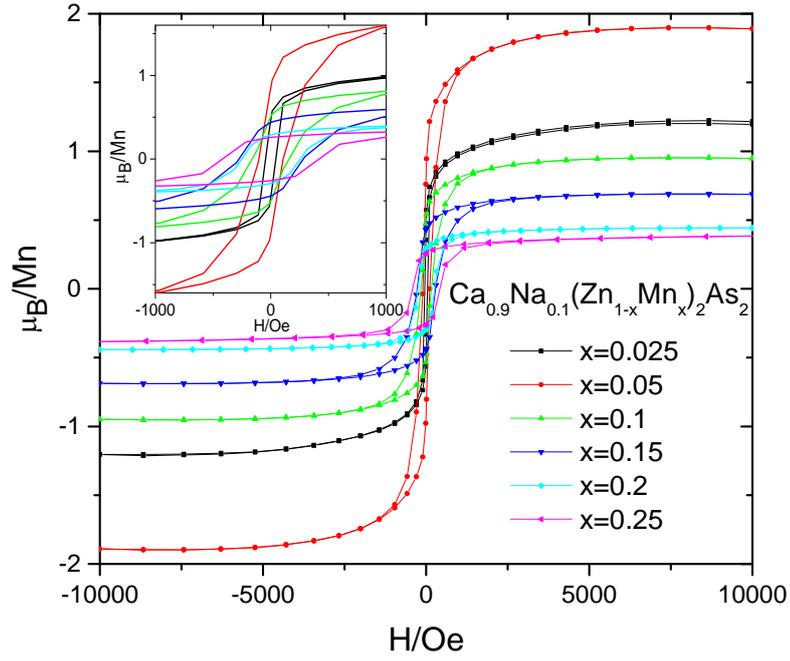

c

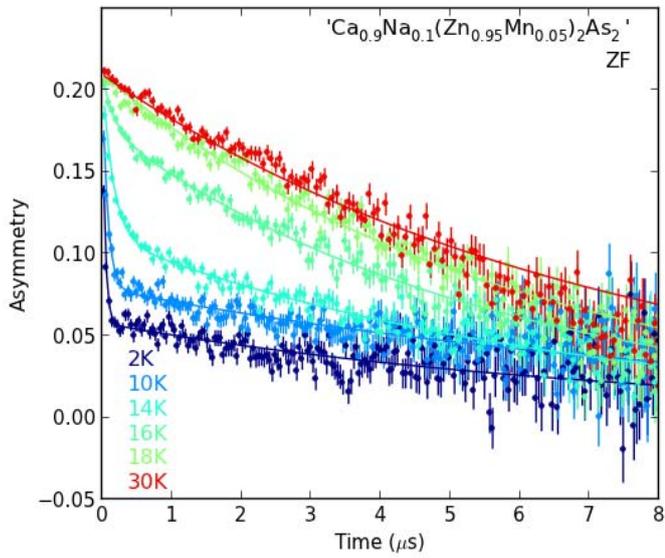

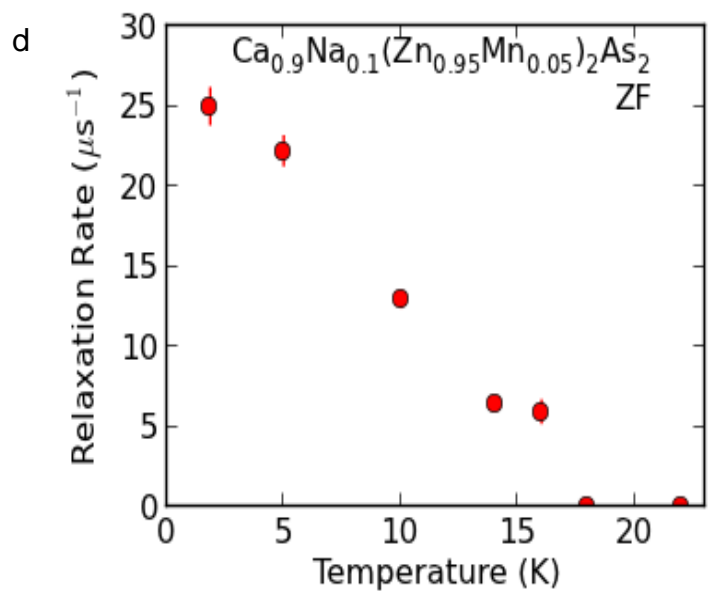

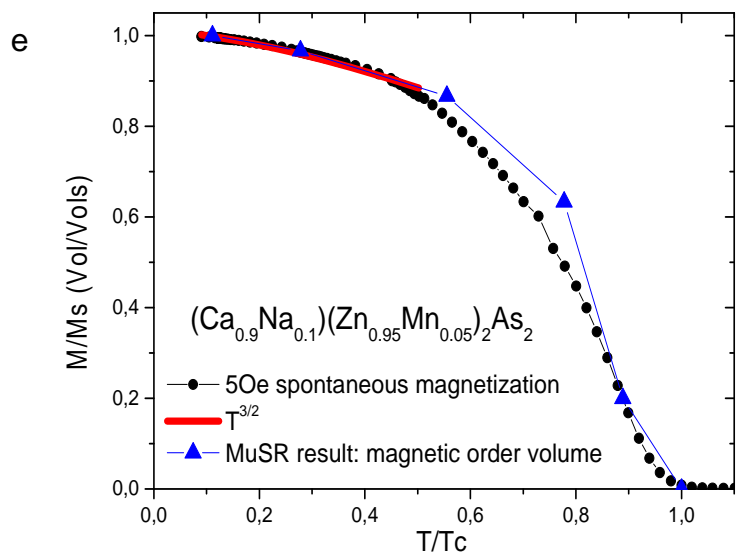

**Fig.3**

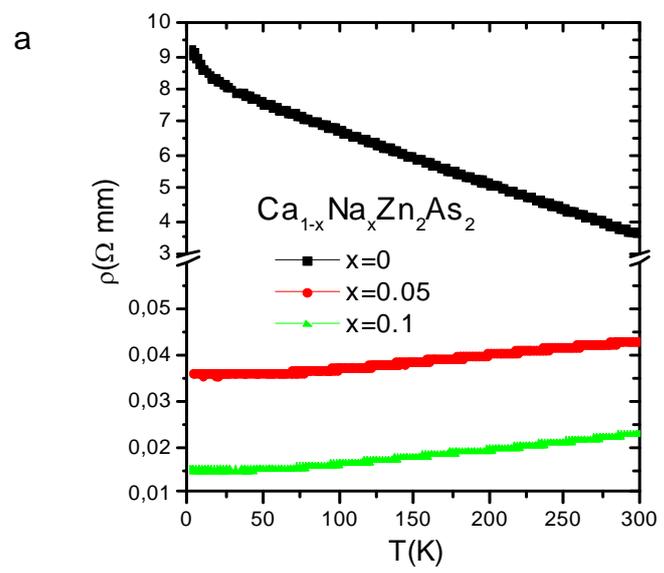

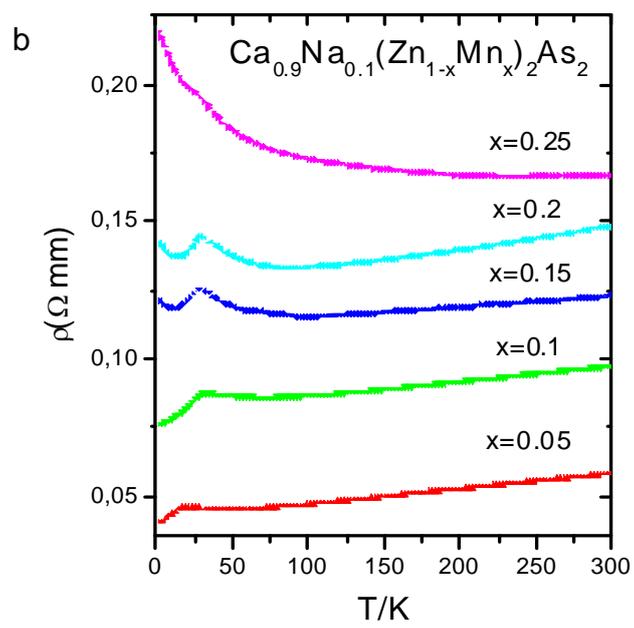

c

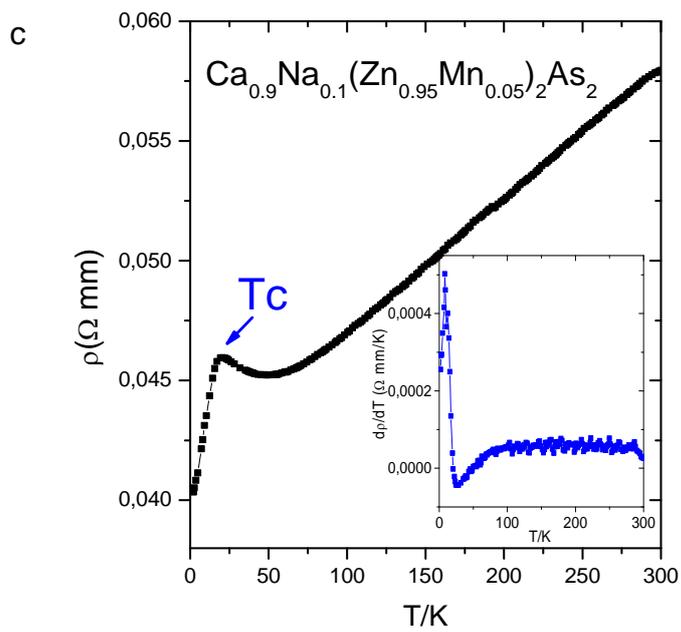

d

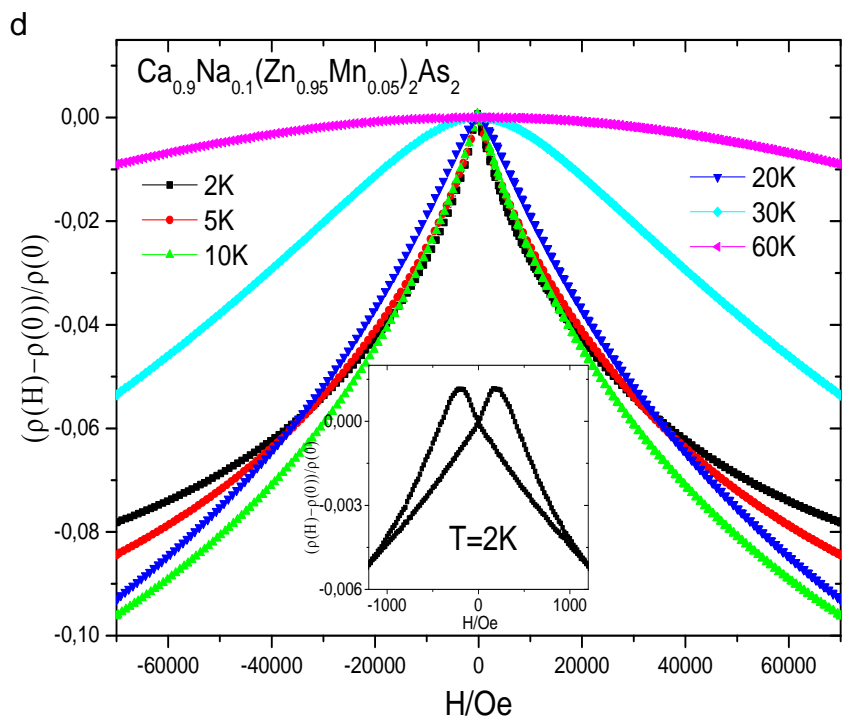

e

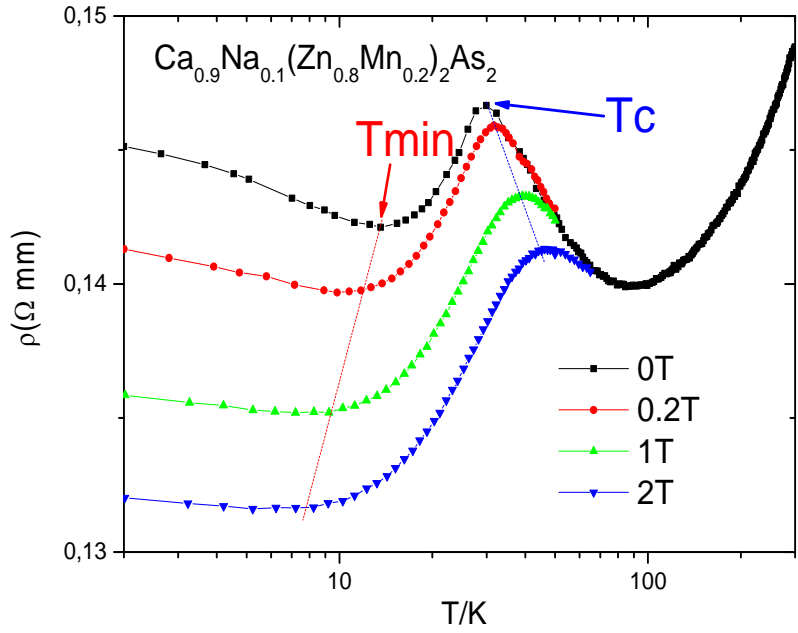

f

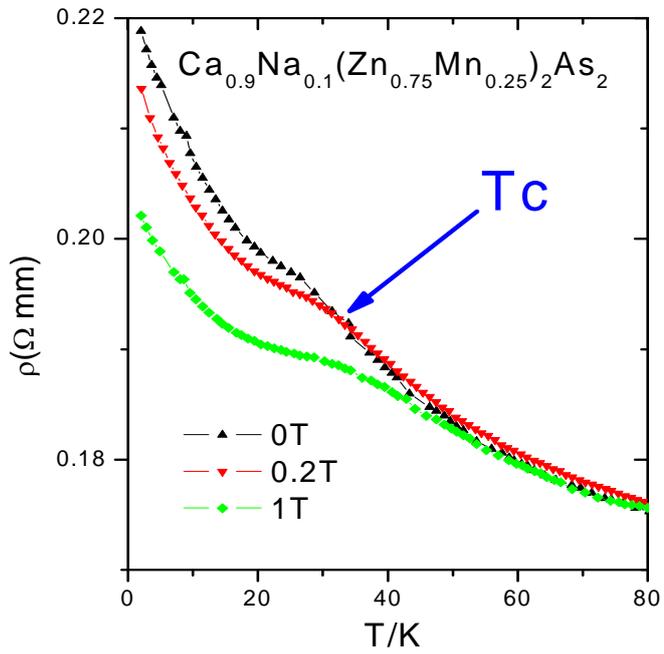

**Fig. 4**

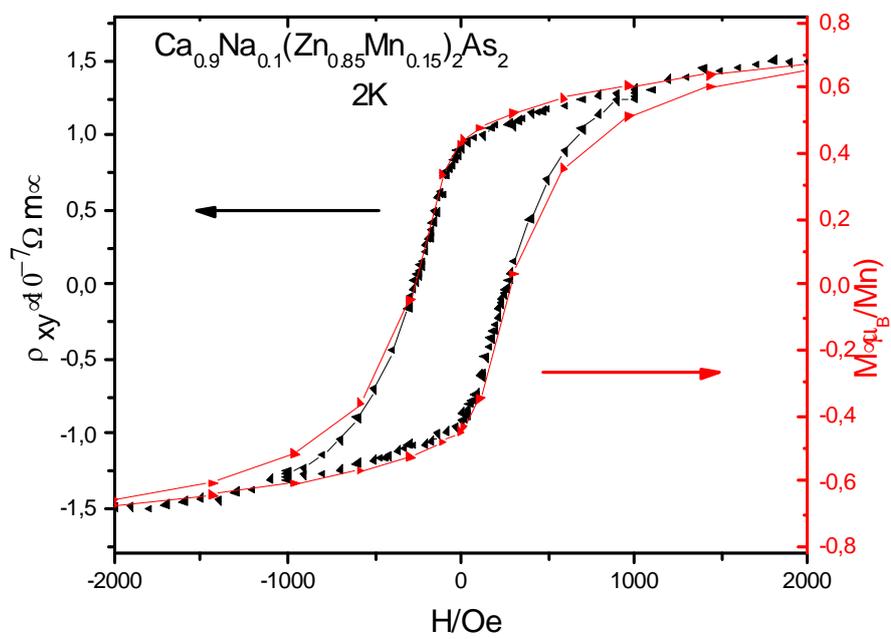